\documentclass [12 pt]{article}
\def\be{\begin{equation}}
\def\eea{\end{eqnarray}}
\def\bea{\begin{eqnarray}}
\def\ee{\end{equation}}

\usepackage[psamsfonts]{amssymb}
\usepackage{amsmath}
\author{F. Kheirandish$^{1}$ \footnote{fardin$_{-}$kh@phys.ui.ac.ir} and M.
Amooshahi$^{1}$ \footnote{amooshahi@sci.ui.ac.ir}
\\ $^{1}$ {\small Department of Physics, University of Isfahan,}
\\ {\small Hezar Jarib Ave., Isfahan, Iran.}}
\title{Quantum electromdynamics in a linear absorbing dielectric medium }
\begin{document}
\maketitle
\begin{abstract}
The eletromagnetic field in a linear absorptive dielectric medium,
is quantized in the framework of the damped polarization model. A
Hamiltonian containing a reservoir with continuous degrees of
freedom, is proposed. The reservoir minimally interacts with the
dielectric polarization and the electromagnetic field. The
Lagevin-Schrodinger equation is obtained as the equation of
motion of the polarization field. The radiation reaction
electromagnetic field is considered. For a homogeneous medium, the
equations of motion are solved using the Laplace transformation
method.
\end{abstract}
\section{Introduction}
In a homogeneous and nondispersive medium, the photon is
associated with only the transverse cmponents of the
electromagnetic field, in contrast in an inhomogeneous
nondispersive medium, the transverse and the longitudinal degrees
of freedom are coupled. In this case the quantization of the
electromagnetic field can be accomplished by employing a
generalized gauge [1,2], that is,
$\vec{\nabla}.(\varepsilon(\vec{r})\vec{A})=0$, where
$\varepsilon(\vec{r})$ is the space dependent dielectric
function. Generalization of this quantization to an anisotropic
nondispersive medium is straightforward by using of the gauge
$\sum_{j=1}^3\frac{\partial}{\partial x_j}(\varepsilon_{ij}(\vec{r})\vec{A}_j)=0$ [3].\\
The quantization in a dispersive and absorptive dielectric,
represents one of the most and interesting problems in quantum
optics, because it gives a rigorous test of our understanding of
the interaction of light with matter. The dissipative nature of a
medium is an immediate consequence of its disspersive character
and vice versa according to the Kramers-Kronig relations. This
means that the validities of the electromagnetic quantization in
nondissipative but dispersive media is restricted to some range of
frequencies for which the imaginary part of frequency dependent
dielectric function is negligible, otherwise there will be
inconsistencies.\\
In the scheme of Lenac [4], for dispersive and nonabsorptive
dielectric media starting with the fundamental equations of motion
, the medium is described by a dielectric function $
\varepsilon(\vec{r},\omega)$, without any restriction on its
spatial behavior. In this scheme it is assumed that there is no
losses in the system, so the dielectric function is real for the
whole space. The procedure is based on an expansion of the total
field in terms of the coupled eigenmodes, orthogonality relations
are
derived and equal-time commutation relations are discussed.\\
Huttner and Barnett [5] have presented a canonical quantization
for the electromagnetic field inside the dispersive and absorptive
dieletrics based on a microscopic model in which the medium is
represented by a collection of the interacting matter fields. The
absorptive character of the medium is modeled through the
interaction of the matter fields with a reservoir consisting of a
continuum of the Klein-Gordon fields. In their model,
eigen-operators for the coupled systems are calculated and the
electromagnetic field has been expressed in terms of them, the
dielectric function is derived and it is shown to satisfy the Kramers-Kronig relations.\\
Matloob [6] has quantized the electromagnetic field in a linear
isotropic medium by associating a damped harmonic oscillator with
each mode of the radiation field. A canonical approach has been
used to quantizing a damped quantum oscillator. The conjugate
momentum is defined and a quantum mechanical Hamiltonian is introduced.\\
Gruner and Welsh [7] have given a quantization scheme for the
radiation field in dispersive and absorptive linear dielectric by
starting from the phenomenological Maxwell equations, there the
properties of the dielectric are described by a permitivity
consistent with the Kramers-Kronig relations, an expansion of the
field operators is performed based on the Green function of the
classical Maxwell equations which preserves the equal-time
canonical commutation relations. In particular, in frequency
intervals with approximately vanishing absorption, the concept of
quantization through mode expansion for dispersive dielectrics is
recognized. The theory further reveals that weak absorption gives
rise to space-dependent mode operators which spatially evolve
according to the quantum Langevin equations in the space domain.

Suttorp and Wubs [8] in the framework of a damped polarization
model, have quantized the electromagnetic field in an absorptive
medium with spatial dependendence of its parameters. They have
solved equations of motion of the dielectric polarization and the
eletromagnetic field by means of the Laplace transformation for
positive and negative times. The operators that diagonalize the
Hamiltonian are found as linear combinations of canonical
variables with coefficients depending on the electric
susceptibility and the dielectric Green function. The time
dependence of the electromagnetic field and dielectric
polarization are determined.
 Present authors [9] have given a
model in which a linear absorptive dielectric is modeled by two
independent reservoirs which are called electrical and magnetic
reservoirs. In this model, the electrical and magnetic
polarization densities, for the reservoirs, are defined. The
electrical and magnetic polarization densities interact with the
displacement vector field and the magnetic field respectively.
Both structural and Maxwell
equations are obtained as the Heisenberg equations of motion.\\
 There are some another approaches for quantizing the electromagnetic field,
 see for example [10-19].\\
In this paper in the framework of the damped polarization model,
we introduce a method in which the electromagnetic and dielectric
polarization fields, interact with a reservoir through a minimal
coupling term. We obtain the Langevin-Schrodinger equation as the
equation of motion of the dielectric polarization and calculate
the radiation reaction electromagnetic field in terms of the
polarization field. Finally we solve the equations of motion in a
homogeneous dielectric using the Laplace transformation.
\section{Quantum dynamics}
We take a model in which the polarization of the dielectric and
the electromagnetic field, as quantum fields, interact with a
reservoir through a minimal coupling iterm. In this model we can
obtain the Maxwell equations and the Langevin-Schrodinger
equation as the equations of motion of the polarization density
of the dielectric.
 The vector potential for the electromagnetic field
in the coulomb gauge, $\nabla.\vec{A}=0$, can be expanded in terms
of the plane waves
\begin{equation}\label{QEDMD1}
\vec{A}(\vec{r},t)=\sum_{\lambda=1}^2\int
d^3\vec{k}\sqrt{\frac{\hbar}{2(2\pi)^3\varepsilon_0\omega_{\vec{k}}}}
[a_{\vec{k}\lambda}(t)e^{i\vec{k}\cdot\vec{r}}+a_{\vec{k}\lambda}^\dag(t)
e^{-i\vec{k}\cdot\vec{r}}]\vec{e}(\vec{k},\lambda),
\end{equation}
where $\omega_{\vec{k}}=c|\vec{k}|$ and $\varepsilon_0$ is
transitivity of vacuum, $\vec{e}(\vec{k},\lambda)$ are
polarization unit vectors
\begin{eqnarray}\label{QEDMD2}
&&\vec{e}(\vec{k},\lambda)\cdot\vec{e}(\vec{k},\lambda')=
\delta_{\lambda\lambda'},\nonumber\\
&&\vec{k}\cdot\vec{e}(\vec{k},\lambda)=0,\hspace{1.50
cm}\lambda=1,2,
\end{eqnarray}
Canonical momentum density of the electromagnetic field is defined
by
\begin{equation}\label{QEDMD3}
\vec{\pi}_F(\vec{r},t)=-i\varepsilon_0\sum_{\lambda=1}^2\int
d^3\vec{k}\sqrt{\frac{\hbar\omega_{\vec{k}}}{2(2\pi)^3\varepsilon_0}}
[a_{\vec{k}\lambda}(t)e^{i\vec{k}\cdot\vec{r}}-a_{\vec{k}\lambda}^\dag(t)e^{-i\vec{k}\cdot
\vec{r}}]\vec{e}(\vec{k},\lambda).
\end{equation}
The creation and annihilation operators $
a_{\vec{k}\lambda}^\dag$, $ a_{\vec{k}\lambda}$, in any instant of
time, satisfy the comutation relations
\begin{equation}\label{QEDMD4}
[a_{\vec{k}\lambda}(t),a_{\vec{k'}\lambda'}^\dag(t)]=\delta_{\lambda\lambda'}
\delta(\vec{k}-\vec{k'}),
\end{equation}
and their time dependence is to be determined from the total
Hamiltonian. Comutation relations (\ref{QEDMD4}) lead to
\begin{equation}\label{QEDMD5}
[\vec{A}_i( \vec{r},t),\pi_{Fj}(
\vec{r'},t)]=i\hbar\delta_{ij}^\bot(\vec{r}-\vec{r'}),
\end{equation}
where $\delta_{ij}^\bot(\vec{r}-\vec{r'})=\frac{1}{(2\pi)^3}\int
d^3\vec{k}e^{i\vec{k}.(\vec{r}-\vec{r'})}(\delta_{ij}-\frac{k_ik_j}{|\vec{k}|^2})
$, is the transverse delta function. The Hamiltonian of the
quantum electromagnetic field is
\begin{equation}\label{QEDMD6}
H_F=\int d^3\vec{r}[\frac{\pi_F^2}{2\varepsilon_0}+\frac{(
\nabla\times\vec{A})^2}{2\mu_0}]=\sum_{\lambda=1}^2\int
d^3\vec{k}\hbar\omega_{\vec{k}}(a_{\vec{k}\lambda}^\dag
a_{\vec{k}\lambda}+\frac{1}{2}).
\end{equation}
The existence of electrical field inside a dielectric medium,
cause electrical charges to move from their stable positions. Let
$\alpha(\vec{r})$ be the density of the displaced charges and $
\vec{Y}(\vec{r},t)$ be their displacement, then the polarization
density induced in the medium is $ \vec{P}=\alpha\vec{Y}$. We can
take $ \vec{Y} $ as a quantum field and expand it in terms of the
plane waves
\begin{equation}\label{QEDMD7}
\vec{Y}(\vec{r},t)=\sum_{\nu=1}^3\int\frac{d^3\vec{k}}{\sqrt{2(2\pi)^3}}
[d_{\vec{k}\nu}(t)e^{i\vec{k}.\vec{r}}+d_{\vec{k}\nu}^\dag(t)e^{-i\vec{k}.
\vec{r}}]\vec{u}_\nu(\vec{k}),
\end{equation}
where $ \vec{u}_\nu(\vec{k})=\vec{e}(\vec{k},\nu) $ for $ \nu=1,2
$ and $\vec{u}_3(\vec{k})=\hat{k}=\frac{\vec{k}}{|\vec{k}|}$. The
sum over $\nu=1,2$, gives the transverse component of $\vec{Y}$
and $\nu=3$, is the longitudinal component. We can write the
canonical momentum density of the quantum field $\vec{Y}$ as
\begin{equation}\label{QEDMD8}
\vec{\pi}_Y(\vec{r},t)=-i\sum_{\nu=1}^3\int
d^3\vec{k}\sqrt{\frac{\hbar}{2(2\pi)^3}}[d_{\vec{k}\nu}(t)e^{i\vec{k}.
\vec{r}}-d_{\vec{k}\nu}^\dag(t)e^{-i\vec{k}.\vec{r}}]\vec{u}_\nu(\vec{k}).
\end{equation}
Imposing the commutation relations
\begin{equation}\label{QEDMD9}
[d_{\vec{k}\nu}(t),d_{\vec{k'}\nu'}^\dag(t)]=\delta_{\nu\nu'}\delta(\vec{k}-\vec{k'}),
\end{equation}
on $d_{\vec{k}\nu}$ and $d_{\vec{k}\nu}^\dag$, leads to the
following commutation relations for vector fields $\vec{Y}$ and
$\vec{\pi}_Y$
\begin{equation}\label{QEDMD10}
[\vec{Y}_i(\vec{r},t),\vec{\pi}_{Yj}(\vec{r'},t)]=i\hbar\delta_{ij}\delta(\vec{r}-\vec{r'}).
\end{equation}
Let us assume a spring force $
-\rho(\vec{r})\omega_0^2(\vec{r})\vec{Y}$, exerted on an element
of the medium with volume $d^3\vec{r}$ centered in $\vec{r}$,
where $\rho(\vec{r})$, is the mass density of the displaced
charges, then the Hamiltonian of the quantum field $\vec{Y}$ can
be written as
\begin{equation}\label{QEDMD11}
H_Y=\int
d^3\vec{r}[\frac{\vec{\pi}_Y^2}{2\rho}+\frac{1}{2}\rho(\vec{r})\omega_0^2(\vec{r})\vec{Y}^2
].
\end{equation}
If the damping forces together with the electrical force are
exerted on the elements of the dielectric, then in the damped
polarization model, it will be assumed that the damping forces are
caused by a quantum field, which is called the reservoir or the
environment. In this model, the equation of motion of the quantum
field $\vec{Y}$ and the reservoir together with the Maxwell
equations, can be obtained from the Heisenberg equations using the
total Hamiltonian
\begin{equation}\label{QEDMD12}
H=\int
d^3\vec{r}[\frac{(\vec{\pi}_Y-\alpha\vec{A}-\vec{R})^2}{2\rho}+
\frac{1}{2}\rho(\vec{r})\omega_0^2(\vec{r})\vec{Y}^2+\frac{(\alpha\vec{Y})\cdot
(\alpha\vec{Y})^\|}{2\varepsilon_0}]+H_F+H_R,
\end{equation}
where $(\alpha\vec{Y})^\|$, is the longitudinal component of the
polarization density and $ H_R $ is the Hamiltonian of the
reservoir or environment defined as
\begin{equation}\label{QEDMD13}
H_R=\sum_{\nu=1}^3\int d^3\vec{k}\int
d^3\vec{q}\hbar\omega_{\vec{k}}[
b_\nu^\dag(\vec{k},\vec{q},t)b_\nu(\vec{k},\vec{q},t)+\frac{1}{2}],
\end{equation}
with the commutation relations
\begin{equation}\label{QEDMD14}
[b_\nu(\vec{k},\vec{q},t),b_{\nu'}^\dag(\vec{k'},\vec{q'},t)]=
\delta_{\nu\nu'}\delta(\vec{k}-\vec{k'})\delta(\vec{q}-\vec{q'}).
\end{equation}
The operator $\vec{R}(\vec{r},t)$, plays a crucial rule in the
interaction between the reservoir with the electromagnetic field
and the quantum field $\vec{Y}$, by definition
\begin{equation}\label{QEDMD15}
\vec{R}=\sum_{\nu=1}^3\int d^3\vec{k}\int
\frac{d^3\vec{q}}{\sqrt{(2\pi)^3}}[f(
\omega_{\vec{k}},\vec{r})b_\nu(\vec{k},\vec{q},t)e^{i\vec{q}\cdot\vec{r}}+f^*(
\omega_{\vec{k}},\vec{r})b_\nu^\dag
(\vec{k},\vec{q},t)e^{-i\vec{q}\cdot\vec{r}}]\vec{u}_\nu(\vec{q}).
\end{equation}
The function $f(\omega_{\vec{k}},\vec{r})$, is called a coupling
function which is position dependent in an inhomogeneous medium.
 According to the total Hamiltonian (\ref{QEDMD12}), the reservoir
interacts with both the electromagnetic field and the quantum
field $\vec{Y}$, while in previous models, the reservoir was
interacting only with the field of the polarization density of the dielectric.\\
\subsection{The equation of motion of the vector field $\vec{Y}$ }
Because the damping forces are exerted on the elements of the
dielectric, the quantum field $\vec{Y}$, describes a dissipative
quantum system. Such a system can be described in terms of the
Langevin equation [20] which has a broad and general application.
This description can be formulated using a coupling between the
system and a quantum mechanical heat-bath. The Heisenberg equation
for the dissipative system, takes the form of
Langevin-Schrodinger equation. The equation of motion of the
position operator $\vec{x}(t)$, is
\begin{equation}
m\ddot{\vec{x}}+\int_0^t dt'\mu(t-t')\dot{\vec{x}}
(t')+V(\vec{x})=\vec{F}_N(t),
\end{equation}
where the coupling with the heat-bath is described by two terms,
a term that describes the absorption of energy by the heat-bath
which is characterized by the memory function $\mu(t)$ and a
fluctuating term, characterized by the operator valued noise
force $\vec{F}_N(t)$. Both terms are necessary for a consistent
description of a quantum damped system.
 In the present model, with the
proposed Hamiltonian (\ref{QEDMD12}), we can obtain the
Langevin-Schrodinger equation for the quantum field $\vec{Y}$ by
 combining the Heisenberg equations of the
reservoir and the field $\vec{Y}$.
  The equations of motion for the fields $\vec{Y}$ and $\vec{\pi}_Y $, can be obtained from
the Heisenberg equations
\begin{eqnarray}\label{QEDMD16}
&&\frac{\partial\vec{Y}}{\partial
t}=\frac{i}{\hbar}[H,\vec{Y}]=\frac{\vec{\pi}_Y-\alpha\vec{A}-\vec{R}}{\rho},\nonumber\\
&&\frac{\partial\vec{\pi}_Y}{\partial
t}=\frac{i}{\hbar}[H,\vec{\pi}_Y]=-\rho\omega_0^2(\vec{r})\vec{Y}-\frac{\alpha}
{\varepsilon_0}(\alpha\vec{Y})^\|,
\end{eqnarray}
so
\begin{equation}\label{QEDMD17}
\rho\frac{\partial^2\vec{Y}}{\partial
t^2}+\rho\omega_0^2(\vec{r})\vec{Y}=-\alpha\frac{\partial\vec{A}}{\partial
t}-\frac{\alpha}{\varepsilon_0}(\alpha\vec{Y})^\|-\frac{\partial\vec{R}}{\partial
t}=\alpha\vec{E}^\bot+\alpha\vec{E}^\|-\frac{\partial\vec{R}}{\partial
t},
\end{equation}
where $\vec{E}^\bot=-\frac{\partial \vec{A}}{\partial t}$  and $
\vec{E}^\|=-\frac{(\alpha\vec{Y})^\|}{\varepsilon_0} $, are
transverse and longitudinal components of the electrical field
respectively. Using (\ref{QEDMD14}), the Heisenberg equation for
$ b_\nu(\vec{k},\vec{q},t)$ can be obtained as
\begin{eqnarray}\label{QEDMD18}
&&\dot{b}_\nu(\vec{k},\vec{q},t) =\frac{i}{\hbar}[H,
b_\nu(\vec{k},\vec{q},t)]=-i\omega_{\vec{k}}b_\nu(\vec{k},\vec{q},t)+\nonumber\\
&&+\frac{i}{\hbar\sqrt{(2\pi)^3}}\int
d^3\vec{r'}f^*(\omega_{\vec{k}},\vec{r'})e^{-i\vec{q}\cdot\vec{r'}}
\frac{\partial\vec{Y}(\vec{r'},t)}{\partial
t}\cdot\vec{u}_\nu(\vec{q}),
\end{eqnarray}
with the formal solution
\begin{eqnarray}\label{QEDMD19}
&&{b}_\nu(\vec{k},\vec{q},t) =e^{-i\omega_{\vec{k}}t}b_\nu(\vec{k},\vec{q},0)+\nonumber\\
&&+\frac{i}{\hbar\sqrt{(2\pi)^3}}\int_0^t
dt'e^{-i\omega_{\vec{k}}(t-t')}\int
d^3\vec{r'}f^*(\omega_{\vec{k}},\vec{r'})e^{-i\vec{q}\cdot\vec{r'}}
\frac{\partial\vec{Y}(\vec{r'},t')}{\partial
t'}\cdot\vec{u}_\nu(\vec{q}).
\end{eqnarray}
Substituting $ {b}_\nu(\vec{k},\vec{q},t)$, from (\ref{QEDMD19})
in (\ref{QEDMD17}), gives
\begin{eqnarray}\label{QEDMD20}
&&\rho\ddot{\vec{Y}}+\rho\omega_0^2(\vec{r})\vec{Y}+\int_0^t d
t'\gamma(t-t',\vec{r})\dot{\vec{Y}}(\vec{r},t')=\alpha\vec{E}^\bot(\vec{r},t)+
\alpha\vec{E}^\|(\vec{r},t)+\vec{\xi}(\vec{r},t),\nonumber\\
&&
\end{eqnarray}
where
\begin{equation}\label{QEDMD21}
\gamma(t-t',\vec{r})=\frac{8\pi}{\hbar c^3}\int_0^\infty d\omega
\omega^3|f(\vec{r},\omega)|^2\cos\omega(t-t'),
\end{equation}
is the memory function and describes the absorption of energy of
the medium by the reservoir. The field $\vec{\xi}(\vec{r},t)$
\begin{eqnarray}\label{QEDMD22}
&&\vec{\xi}(\vec{r},t)=i\sum_{\nu=1}^3\int
d^3\vec{k}\int\frac{d^3\vec{q}}{\sqrt{(2\pi)^3}}\omega_{\vec{k}}
[f(\vec{r},\omega_{\vec{k}})b_\nu(\vec{k},\vec{q},0)e^{-i\omega_{\vec{k}}t+
i\vec{q}\cdot\vec{r}}-\nonumber\\
&&f^*(\vec{r},\omega_{\vec{k}})b_\nu^\dag(\vec{k},\vec{q},0)
e^{+i\omega_{\vec{k}}t-i\vec{q}\cdot\vec{r}}]\vec{u}_\nu(\vec{q}).
\end{eqnarray}
is the noise field associated with the absorption and has a zero
expectation value in the eigenstates of the reservoir.
 The equation (\ref{QEDMD20}), is called the
Langevin-Schrodinger equation for the damped quantum field
$\vec{Y}$. The term $ \int_0^t d
t'\gamma(t-t',\vec{r})\dot{\vec{Y}}(\vec{r},t')$, usually
generates a damping force for $\vec{Y}$, for example if we take
\begin{equation}\label{QEDMD23}
|f(\omega)|^2=\frac{\beta c^3\hbar}{4\pi^2\omega^3},
\end{equation}
 we obtain
\begin{equation}\label{QEDMD24}
\rho\ddot{\vec{Y}}+\rho\omega_0^2(\vec{r})\vec{Y}+
\beta\dot{\vec{Y}}(\vec{r},t)=\alpha\vec{E}^\bot(\vec{r},t)+\alpha\vec{E}^\|
(\vec{r},t)+\vec{\tilde{\xi}}(\vec{r},t),
\end{equation}
which has a damping term proportional to the velocity and
$\vec{\tilde{\xi}}(\vec{r},t)$, is the noise field
(\ref{QEDMD22}) with the coupling function (\ref{QEDMD23}).
\subsection{The radiation reaction}
In QED, a charged particle in quantum vacuum interacts with the
vacuum field and its own field known as the radiation reaction. In
classical electrodynamics, there is only the radiation reaction
field that acts on a charged particle in the vacuum. The vacuum
and radiation reaction fields have a fluctuation-dissipation
connection and both are required for the consistency of QED. For
example the stability of the ground state, atomic transitions and
lamb shift can only be explained by taking into account both
fields. If self reaction was alone the atomic ground state would
not be stable. When a quantum mechanical system interacts with
the  vacuum quantum field, the coupled Heisenberg equations for
both the system and the quantum vacuum field give us the radiation
reaction field, for example it can be shown that the radiation
reaction for a charged harmonic oscillator is $ \frac{2e^2}{3c^3}
$ [21].
 One can find the Heisenberg equation for the annihilation
operators $a_{\vec{k}\lambda}$
\begin{eqnarray}\label{QEDMD25}
&&\dot{a}_{\vec{k}\lambda}=\frac{i}{\hbar}[H,a_{\vec{k}\lambda}]=-i\omega_{\vec{k}}
a_{\vec{k}\lambda}+i\int
d^3\vec{r'}\dot{\vec{Y}}(\vec{r'},t)\cdot\vec{e}(\vec{k},\lambda)\frac{\alpha
e^{-i\vec{k}\cdot\vec{r'}}}{\sqrt{2(2\pi)^3\varepsilon_0\hbar\omega_{\vec{k}}}},\nonumber\\
&&
\end{eqnarray}
with the formal solution
\begin{eqnarray}\label{QEDMD25}
&&{a}_{\vec{k}\lambda}(t)=e^{-i\omega_{\vec{k}}t}a_{\vec{k}\lambda}(0)
+i\int_0^t d t'e^{-i\omega_{\vec{k}}(t-t')}\int
d^3\vec{r'}\dot{\vec{Y}}(\vec{r'},t')\cdot\vec{e}(\vec{k},\lambda)\frac{\alpha
e^{-i\vec{k}\cdot\vec{r'}}}{\sqrt{2(2\pi)^3\varepsilon_0\hbar\omega_{\vec{k}}}}.\nonumber\\
&&
\end{eqnarray}
Substituting  ${a}_{\vec{k}\lambda}(t)$ from (\ref{QEDMD25}) in $
\vec{E}^\bot=-\frac{\partial\vec{A}}{\partial t}$, we obtain
\begin{equation}\label{QEDMD26}
\vec{E}^\bot=\vec{E}_0^\bot+\vec{E}_{RR}^\bot,
\end{equation}
where
\begin{equation}\label{QEDMD27}
\vec{E}_0^\bot=i\int d^3\vec{k}\sqrt
{\frac{\hbar\omega_{\vec{k}}}{2(2\pi)^3\varepsilon_0}}[a_{\vec{k}\lambda}(0)e^{-i\omega_{\vec{k}}
t+i\vec{k}\cdot\vec{r}}-a_{\vec{k}\lambda}^\dag(0)e^{+i\omega_{\vec{k}}
t-i\vec{k}\cdot\vec{r}}]\vec{e}(\vec{k},\lambda),
\end{equation}
is the transverse vacuum field and
\bea\label{QEDMD28}
\vec{E}_{RR}^\bot&=&-\frac{1}{(2\pi)^3\varepsilon_0}\int
d^3\vec{r'}\alpha(\vec{r'})\int d^3\vec{k}\int_0^t
dt'\cos[\omega_{\vec{k}}(t-t')+\vec{k}\cdot(\vec{r}-\vec{r'})]\nonumber\\
&\times&\{\dot{\vec{Y}}(\vec{r'},t')-[\hat{k}.\dot{\vec{Y}}(\vec{r'},t')]\hat{k}\}
\eea
is the transverse radiation reaction electrical field [21].
\subsection{The Maxwell equations}
The Maxwell equations can be obtained as the Heisenberg equations
of $ \vec{A} $ and $ \vec{\pi}_F $
\begin{eqnarray}\label{QEDMD29}
&&\frac{\partial\vec{A}}{\partial
t}=\frac{i}{\hbar}[H,\vec{A}]=\frac{\vec{\pi}_F}{\varepsilon_0},\nonumber\\
&&\frac{\partial\vec{\pi}_F}{\partial
t}=\frac{i}{\hbar}[H,\vec{\pi}_F]=(\alpha
\dot{\vec{Y}})^\bot-\frac{\nabla\times\nabla\times\vec{A}}{\mu_0},
\end{eqnarray}
after eliminating $ \vec{\pi}_F $,
\begin{equation}\label{QEDMD30}
\varepsilon_0\frac{\partial^2\vec{A}}{\partial t^2}=(\alpha
\dot{\vec{Y}})^\bot-\frac{\nabla\times\nabla\times\vec{A}}{\mu_0}.
\end{equation}
By defining the displacement vector field $ \vec{D} $ as $
\vec{D}=\vec{D}^\bot=(\alpha\vec{Y})^\bot+\varepsilon_0\vec{E}^\bot
$ and $ \vec{B}=\nabla\times\vec{A} $, the equation
(\ref{QEDMD30}) can be rewritten as
\begin{equation}\label{QEDMD31}
\nabla\times\vec{B}=\mu_0\frac{\partial\vec{D}^\bot}{\partial t}
\end{equation}
which is the familiar form of the Maxwell equation in a
non-magnetic dielectric medium. In the absence of the external
electrical charge density, we have $
\vec{D}^\|=\varepsilon_0\vec{E}^\|+(\alpha\vec{Y})^\|=0 $ or $
\vec{E}^\|=-\frac{(\alpha\vec{Y})^\|}{\varepsilon_0} $.
\section{Solution of the Heisenberg equations}
The equation of motion defined in the previous sections, can be
solved by using the Laplace transformation method. For any time
dependent operator $g(t)$, the forward and the backward
$(g^{f(b)}(t))$, Laplace transformation, are defined respectively
as
\bea\label{QEDMD32}
g^f(s)&=&\int_0^\infty dt
g(t)e^{-st},\nonumber\\
g^b(s)&=&\int_0^\infty dt g(-t)e^{-st}. \eea Now taking the
Laplace transformation of the equation (\ref{QEDMD20})
\begin{eqnarray}\label{QEDMD34}
&&\alpha\vec{Y}^f(\vec{r},s)=\alpha\vec{Y}_N^f(\vec{r},s)+\varepsilon_0
\tilde{\chi}(\vec{r},s)\vec{E}^f(\vec{r},s),\nonumber\\
&&\alpha\vec{Y}^b(\vec{r},s)=\alpha\vec{Y}_N^b(\vec{r},s)+\varepsilon_0
\tilde{\chi}(\vec{r},s)\vec{E}^b(\vec{r},s),
\end{eqnarray}
where
\begin{equation}\label{QEDMD35}
\tilde{\chi}(\vec{r},s)=\frac{\alpha^2(\vec{r})}{\varepsilon_0[\rho(\vec{r})
s^2+\rho(\vec{r})\omega_0^2(\vec{r})+s\tilde{\gamma}(\vec{r},s)]},\hspace{0.4
cm}\tilde{\gamma}(\vec{r},s)=8\pi s\int_0^\infty
d\omega\frac{\omega^3|f(\vec{r},\omega|^2}{s^2+\omega^2}.
\end{equation}
The function $\tilde{\chi}(\vec{r},s)$, is the Laplace
transformation of the electrical susceptibility and
\begin{eqnarray}\label{QEDMD36}
\alpha\vec{Y}^f_N(\vec{r},s)&=&\frac{\varepsilon_0}{\alpha}\tilde{\chi}(\vec{r},s)\{\vec{\xi}^f(\vec{r},s)+\rho
s\vec{Y}(\vec{r},0)+\vec{\pi}_Y(\vec{r},0)-\alpha\vec{A}(\vec{r},0)\nonumber\\
&-&\vec{R}(\vec{r},0)+\tilde{\gamma}(\vec{r},s)\vec{Y}(\vec{r},0)\},\nonumber\\
\alpha\vec{Y}_N^b(\vec{r},s)&=&\frac{\varepsilon_0}{\alpha}\tilde{\chi}(\vec{r},s)\{\vec{\xi}^b(\vec{r},s)+\rho
s\vec{Y}(\vec{r},0)-\vec{\pi}_Y(\vec{r},0)+\alpha\vec{A}(\vec{r},0)\nonumber\\
&+&\vec{R}(\vec{r},0)+\tilde{\gamma}(\vec{r},s)\vec{Y}(\vec{r},0)\},
\end{eqnarray}
are the forward and backward Laplace transformations of the noise
polarization densities respectively. Another main equation can be
obtained by taking the time derivative of the equation
(\ref{QEDMD30}) and using the relation $
\vec{\nabla}\times\vec{E}^\|=-\vec{\nabla}\times\frac{(\alpha\vec{Y})^\|}{\varepsilon_0}=0
$
\begin{equation}\label{QEDMD37}
-\mu_0\varepsilon_0\frac{\partial^2\vec{E}}{\partial
t^2}=\mu_0\alpha\frac{\partial^2\vec{Y}}{\partial
t^2}+\vec{\nabla}\times\vec{\nabla}\times\vec{E}.
\end{equation}
Equation (\ref{QEDMD37}), can be written in terms of the Laplace
transformed components
\begin{eqnarray}\label{QEDMD38}
&&\vec{\nabla}\times\vec{\nabla}\times\vec{E}^f(\vec{r},s)+\varepsilon_0\mu_0
s^2\tilde{\varepsilon}(\vec{r},s)\vec{E}^f(\vec{r},s)=\vec{J}^f(\vec{r},s),\nonumber\\
&&\vec{\nabla}\times\vec{\nabla}\times\vec{E}^b(\vec{r},s)+\varepsilon_0\mu_0
s^2\tilde{\varepsilon}(\vec{r},s)\vec{E}^b(\vec{r},s)=\vec{J}^b(\vec{r},s),
\end{eqnarray}
where $\tilde{\varepsilon}(\vec{r},s)=1+\tilde{\chi}(\vec{r},s)$,
is the Laplace transformation of the electrical permeability and
\begin{eqnarray}\label{QEDMD39}
&&\vec{J}^f(\vec{r},s)=\vec{\nabla}\times\vec{\nabla}\times\vec{A}(\vec{r},0)-\nonumber\\
&&\mu_0s\pi_F(\vec{r},0)-\mu_0s(\alpha\vec{Y})^\|(\vec{r},0)+
\mu_0s\alpha\vec{Y}(\vec{r},0)-\mu_0s^2\alpha\vec{Y}_N^f(\vec{r},0),\nonumber\\
&&\vec{J}^b(\vec{r},s)=-\vec{\nabla}\times\vec{\nabla}\times\vec{A}(\vec{r},0)-\nonumber\\
&&\mu_0s\pi_F(\vec{r},0)-\mu_0s(\alpha\vec{Y})^\|(\vec{r},0)+
\mu_0s\alpha\vec{Y}(\vec{r},0)-\mu_0s^2\alpha\vec{Y}_N^b(\vec{r},0),\nonumber\\
&&
\end{eqnarray}
are called the source noise densities. In reference [8], the wave
equations (\ref{QEDMD38}) have been solved using the Green
function method. In the following for simplicity, we solve the
Heisenberg equations (\ref{QEDMD17}) and (\ref{QEDMD29}) for an
homogeneous dielectric, that is when $\rho,\omega_0,\alpha,f $
are independent of the position $\vec{r}$. In the case of an
homogeneous and linear dielectric, in the absence of external
charge density, the polarization charge density is zero and
therefore the vector field $\alpha\vec{Y}$, has only the
transverse components. In this case, using the Laplace
transformation of (\ref{QEDMD29}) and (\ref{QEDMD34})
\begin{eqnarray}\label{QEDMD40}
&&\vec{\nabla}\times\vec{\nabla}\times\vec{A}^f(\vec{r},s)+
\mu_0\varepsilon_0s^2\varepsilon(s)\vec{A}^f(\vec{r},s)
=\mu_0s\alpha\vec{Y}_N^f(\vec{r},s)+
\mu_0\varepsilon_0s\varepsilon(s)\vec{A}(\vec{r},0)\nonumber\\
&&-\mu_0\alpha\vec{Y}(\vec{r},0)+\mu_0\vec{\pi}_F(\vec{r},0),\nonumber\\
&&\vec{\nabla}\times\vec{\nabla}\times\vec{A}^b(\vec{r},s)+
\mu_0\varepsilon_0s^2\varepsilon(s)\vec{A}^b(\vec{r},s)
=-\mu_0s\alpha\vec{Y}_N^b(\vec{r},s)+\mu_0\varepsilon_0s\varepsilon(s)\vec{A}(\vec{r},0)\nonumber\\
&&+\mu_0\alpha\vec{Y}(\vec{r},0)-\mu_0\vec{\pi}_F(\vec{r},0).
\end{eqnarray}
The equations (\ref{QEDMD40}), can be solved easily using the
Fourier transformation
\begin{eqnarray}\label{QEDMD41}
\vec{A}(\vec{r},s)&=&\sum_{\lambda=1}^2\int
d^3\vec{k}\sqrt{\frac{\hbar}{2(2\pi)^3\varepsilon_0\omega_{\vec{k}}}}a_{\vec{k}\lambda}(0)e^{i\vec{k}.\vec{r}}\frac{\mu_0\varepsilon_0(s\mp
i\omega_{\vec{k}})
}{\vec{k}^2+\mu_0\varepsilon_0s^2\varepsilon(s)}\vec{e}(\vec{k},\lambda)\nonumber\\
& \mp &\frac{1}{\alpha}\sum_{\lambda=1}^2\int
d^3\vec{k}\sqrt{\frac{\hbar}{2(2\pi)^3}}d_{\vec{k}\lambda}(0)e^{i\vec{k}\cdot\vec{r}}
\frac{\mu_0\varepsilon_0
(\rho\omega_0^2 \pm is)\tilde{\chi}(s)}{\vec{k}^2+\mu_0\varepsilon_0s^2\varepsilon(s)}
\vec{e}(\vec{k},\lambda)\nonumber\\
&-&\frac{1}{\alpha}\int d^3\vec{k}\int
\frac{d^3\vec{q}}{\sqrt{2(2\pi)^3}}\sum_{\lambda=1}^2f(\omega_{\vec{q}})
b_\lambda(\vec{q},\vec{k},0)e^{i\vec{k}\cdot\vec{r}}\frac{\mu_0\varepsilon_0s
\tilde{\chi}(s)}{\vec{k}^2+\mu_0\varepsilon_0
s^2\varepsilon(s)}\vec{e}(\vec{k},\lambda)\nonumber\\
& \pm &\frac{i}{\alpha}\int d^3\vec{k}\int
\frac{d^3\vec{q}}{\sqrt{2(2\pi)^3}}\sum_{\lambda=1}^2
\frac{\omega_{\vec{q}}f(\omega_{\vec{q}})b_\lambda(\vec{q},\vec{k},0)
e^{i\vec{k}\cdot\vec{r}}}{(s \pm i\omega_{\vec{q}})}
\frac{\mu_0\varepsilon_0s\tilde{\chi}(s)}{\vec{k}^2+
\mu_0\varepsilon_0s^2\varepsilon(s)}\vec{e}(\vec{k},\lambda)\nonumber\\
&+&C.C,
\end{eqnarray}
where the upper and lower signs $(\pm)$, give
$\vec{A}^f(\vec{r},s) $ and $ \vec{A}^b(\vec{r},s)$, respectively.
Now $\vec{A}(\vec{r},t)$ for $ t>0 $ is the inverse Laplace
transformation of $\vec{A}^f(\vec{r},s)$ and $
\vec{A}(\vec{r},-t) $ for $ t>0 $, is the inverse Laplace
transformation of $\vec{A}^b(\vec{r},s)$, that is
\begin{eqnarray}\label{QEDMD42}
&&\vec{A}(\vec{r},t)=SOR[e^{st}\vec{A}^f(\vec{r},s)],\hspace{1.00
cm}t>0,\nonumber\\
&&\vec{A}(\vec{r},t)=SOR[e^{-st}\vec{A}^b(\vec{r},s)],\hspace{1.00
cm}t<0,
\end{eqnarray}
 where $SOR[f(s)]$, means the sum of the residues of a complex function $f$.
 If the residues of the complex
functions $\frac{s\mp i\omega_{\vec{k}}
}{\vec{k}^2+\mu_0\varepsilon_0s^2\varepsilon(s)},\frac{(\rho\omega_0^2
\pm is)\tilde{\chi}(s)}{\vec{k}^2+\mu_0\varepsilon_0
s^2\varepsilon(s)}$ and $
\frac{s\tilde{\chi}(s)}{\vec{k}^2+\mu_0\varepsilon_0s^2\varepsilon(s)}$,
with respect to $s$ have negative real parts, as usually it is
the case for an absorptive dielectric, then we find the following
behaviour in $t\rightarrow\infty$
\bea\label{QEDMD43}
\vec{A}(\vec{r},t)&=&\frac{\mu_0\varepsilon_0}{\alpha}\int
d^3\vec{k}\int
\frac{d^3\vec{q}}{\sqrt{(2\pi)^3}}\sum_{\lambda=1}^2\frac{\omega_{\vec{q}}^2f(\omega_{\vec{q}})b_\lambda(\vec{q},\vec{k},0)\tilde{\chi}(-i\omega_{\vec{q}})}{\vec{k}^2-\mu_0\varepsilon_0\omega_{\vec{q}}^2\varepsilon(-i\omega_{\vec{q}})}
e^{-i\omega_{\vec{q}}t+i\vec{k}\cdot\vec{r}}\vec{e}(\vec{k},\lambda)\nonumber\\
&+&C.C.
\eea
We can find the polarization field $
\vec{Y}(\vec{r},t)$ by substituting $\vec{A}^f(\vec{r},s)$ and $
\vec{A}^b(\vec{r},s)$ from (\ref{QEDMD41}) in (\ref{QEDMD34}) and
then taking the inverse Laplace transformation of $
\vec{Y}^f(\vec{r},s)$ and $\vec{Y}^b(\vec{r},s)$.
\section{Conclusion}
By taking a reservoir with continuous degrees of freedom and a
suitable Hamiltonian, we could give a consistent quantization
scheme for electromagnetic field in an absorptive dielectric
medium, in the frame work of the damped polarization model. In
this model, the damped polarization field, was satisfying the
quantum Langevin equation containing the electrical field as the
source term. Using the Laplace transformation, the quantum
Langevin equation solved and the susceptibility of the dielectric
obtained in terms of the coupling function between the reservoir
and the polarization field. Equations containing a noise term for
the vector potential and the electrical field obtained and solved
which led to explicit forms of the vector potential and the
electrical field.

\end{document}